\documentclass[twocolumn, superscriptaddress, secnumarabic, amssymb, nobibnotes, aps, prl, preprintnumbers, nofootinbib]{revtex4-1}

\RequirePackage[colorlinks=true
,urlcolor=blue
,anchorcolor=blue
,citecolor=blue
,filecolor=blue
,linkcolor=blue
,menucolor=blue
,linktocpage=true
,pdfproducer=medialab
,pdfa=true
]{hyperref}

\usepackage{orcidlink}

\usepackage{graphicx}

\usepackage{subfigure}

\usepackage{float}
\usepackage{amsmath}

\usepackage{amssymb}
\usepackage{lipsum}

\usepackage{lineno}

\usepackage{multirow}
\usepackage{setspace}

\newcommand\blfootnote[1]{
\begingroup
\renewcommand\thefootnote{}\footnote{#1}
\addtocounter{footnote}{-1}
\endgroup
}

\newcommand{\SignificanceNuEFullToy}{5.2}
\newcommand{\SignificanceNuMuFullToy}{5.7}

\newcommand{\BackgroundRandomNuE}{$0.025^{+0.015}_{-0.010}$}
\newcommand{\BackgroundRandomNuMu}{$0.22^{+0.09}_{-0.07}$}

\newcommand{\XsecObsResultNuE}{1.2_{-0.7}^{+0.8}}
\newcommand{\XsecObsResultNuMu}{0.5\pm0.2}

\newcommand{\AlphaResultNuE}{2.4^{+1.8}_{-1.4}} 
\newcommand{\AlphaResultNuMu}{0.9\pm0.4} 

\newcommand{\NncContamiLight}{$0.045 _{-0.005}^{+0.004}$ (flux) $\pm 0.003$ (cross section) $_{-0.024}^{+0.076}$ (others)}
\newcommand{\NncContamiCharm}{$0.008 _{-0.004}^{+0.013}$ (flux) $\pm 0.001$ (cross section) $_{-0.004}^{+0.007}$ (others)}

\begin{document}

\preprint{CERN-EP-2024-079}

\title{First Measurement of the $\nu_e$ and $\nu_\mu$ Interaction Cross Sections at the LHC with FASER's Emulsion Detector}

\author{Roshan Mammen Abraham\,\orcidlink{0000-0003-4678-3808}}
\affiliation{Department of Physics and Astronomy, University of California, Irvine, CA 92697-4575, USA}

\author{John Anders\,\orcidlink{0000-0002-1846-0262}}
\affiliation{CERN, CH-1211 Geneva 23, Switzerland}

\author{Claire Antel\,\orcidlink{0000-0001-9683-0890}}
\affiliation{D\'epartement de Physique Nucl\'eaire et Corpusculaire, University of Geneva, CH-1211 Geneva 4, Switzerland}

\author{Akitaka Ariga\,\orcidlink{0000-0002-6832-2466}}
\affiliation{Albert Einstein Center for Fundamental Physics, Laboratory for High Energy Physics, University of Bern, Sidlerstrasse 5, CH-3012 Bern, Switzerland}
\affiliation{Department of Physics, Chiba University, 1-33 Yayoi-cho Inage-ku, 263-8522 Chiba, Japan}

\author{Tomoko Ariga\,\orcidlink{0000-0001-9880-3562}}
\email[Corresponding author. \\E-mail address: ]{ariga@artsci.kyushu-u.ac.jp}
\affiliation{Kyushu University, Nishi-ku, 819-0395 Fukuoka, Japan}

\author{Jeremy Atkinson\,\orcidlink{0009-0003-3287-2196}} 
\affiliation{Albert Einstein Center for Fundamental Physics, Laboratory for High Energy Physics, University of Bern, Sidlerstrasse 5, CH-3012 Bern, Switzerland}

\author{Florian~U.~Bernlochner\,\orcidlink{0000-0001-8153-2719}} 
\affiliation{Universit\"at Bonn, Regina-Pacis-Weg 3, D-53113 Bonn, Germany}

\author{Tobias Boeckh\,\orcidlink{0009-0000-7721-2114}} 
\affiliation{Universit\"at Bonn, Regina-Pacis-Weg 3, D-53113 Bonn, Germany}

\author{Jamie Boyd\,\orcidlink{0000-0001-7360-0726}} 
\affiliation{CERN, CH-1211 Geneva 23, Switzerland}

\author{Lydia Brenner\,\orcidlink{0000-0001-5350-7081}}
\affiliation{Nikhef National Institute for Subatomic Physics, Science Park 105, 1098 XG Amsterdam, Netherlands}

% ...

\collaboration{FASER Collaboration}
\email[E-mail address: ]{faser-publications@cern.ch}
\noaffiliation

\date{\today}

\begin{abstract} 
The first results of the study of high-energy electron neutrino ($\nu_e$) and muon neutrino ($\nu_{\mu}$) charged-current interactions in the FASER$\nu$ emulsion/tungsten detector of the FASER experiment at the LHC are presented. A 128.8~kg subset of the FASER$\nu$ volume was analysed after exposure to 9.5 fb$^{-1}$ of $\sqrt{s} = 13.6$ TeV $pp$ data. Four (eight) $\nu_e$ ($\nu_{\mu}$) interaction candidate events are observed with a statistical significance of 5.2$\sigma$ (5.7$\sigma$). This is the first direct observation of $\nu_e$  interactions at a particle collider and includes the highest-energy $\nu_e$ and $\nu_{\mu}$ ever detected from an artificial source. The interaction cross section per nucleon $\sigma/E_{\nu}$ is measured over an energy range of 560--1740 GeV (520--1760 GeV) for $\nu_e$ ($\nu_{\mu}$) to be 
$(1.2_{-0.7}^{+0.8}) \times 10^{-38}~\mathrm{cm}^{2}\,\mathrm{GeV}^{-1}$ ($(0.5\pm0.2) \times 10^{-38}~\mathrm{cm}^{2}\,\mathrm{GeV}^{-1}$), consistent with Standard Model predictions. These are the first measurements of neutrino interaction cross sections in those energy ranges.
\end{abstract}

\maketitle

\blfootnote{\vspace{1mm}}
\blfootnote{\copyright 2024 CERN for the benefit of the FASER Collaboration.}
\blfootnote{Reproduction of this article or parts of it is allowed as specified in the CC-BY-4.0 license.}

\section{Introduction}
\label{sec:introduction}

To date, neutrino interaction cross sections have not been measured at energies above 300~GeV for $\nu_e$ and between 400~GeV and 6~TeV for $\nu_{\mu}$. One of the primary physics goals of FASER~\cite{Feng:2017uoz, FASER:2018ceo, FASER:2018bac}, the Forward Search Experiment at CERN's Large Hadron Collider (LHC)~\cite{Evans:2008zzb}, is the study of high-energy neutrinos produced in the LHC's proton-proton ($pp$) collisions using the dedicated FASER$\nu$ detector~\cite{FASER:2019dxq,FASER:2020gpr}. FASER$\nu$ is a tungsten/emulsion detector that is located in front of the FASER spectrometer~\cite{FASER:2022hcn}. With FASER$\nu$, charged particle tracks produced by neutrino interactions in the detector can be reconstructed with sub-micron precision. 
This allows us to identify leptons and measure the energies of electrons and momenta of muons, enabling the identification of electron and muon charged-current (CC) neutrino interactions and the measurement of neutrino interaction cross sections in the currently unexplored TeV energy range. 
In this study, we do not measure the charge of the outgoing charged leptons; charge conjugation is implied, and $\nu_{e/\mu}$ represents the sum of both $\nu_{e/\mu}$ and $\bar{\nu}_{e/\mu}$.

The decays of hadrons originating from LHC $pp$ collisions produce a large number of neutrinos, which are focused along the beam collision axis or line of sight (LOS). Neutrinos close to the LOS are characterised by very high energies up to several TeV, and they therefore have relatively large interaction cross sections. 
Since neutrinos only interact weakly, they are not affected by the 100~m of rock or by the magnetic fields between the collision point and the detector. These, however, substantially reduce the rate of background particles.

The first evidence for neutrino interaction candidates produced at the LHC was reported by the FASER Collaboration in 2021~\cite{FASER:2021mtu}. The first observation of muon neutrinos by means of the FASER electronic detector components was achieved in 2023~\cite{FASER:2023zcr}. The SND@LHC Collaboration~\cite{SHiP:2020sos, SNDLHC:2022ihg} confirmed this observation in Ref.~\cite{SNDLHC:2023pun}, studying a different rapidity region.
Nevertheless to date no electron neutrino has ever been directly detected at the LHC.

Neutrinos interact through $\nu_e + N \rightarrow e^- + X$ ($\nu_e$ CC events), $\nu_{\mu} + N \rightarrow \mu^- + X$ ($\nu_{\mu}$ CC events), $\nu_{\tau} + N \rightarrow \tau^- + X$ ($\nu_{\tau}$ CC events)\footnote{$\nu_{\tau}$ CC events are not considered in this analysis due to the small number of expected events in the analyzed sample, as shown in Table~\ref{tab:eventrate} of Appendix B.}, and $\nu_l + N \rightarrow \nu_l + X$ $(l=e,\mu,\tau)$ (neutral-current (NC) events), where $N$ represents a nucleon in the tungsten target, and $X$ represents interaction products. 
At FASER, the main background to neutrino detection arises from neutral hadrons interacting in the detector. These neutral hadrons are generally lower in energy than the neutrino signal, and they can therefore be suppressed with appropriate selections on topological and kinematic variables related to the reconstructed interaction vertices. The additional requirement of a high-energy electron or muon signature further suppresses this background and allows us to distinguish $\nu_{e}$ and $\nu_{\mu}$ CC interactions. 

The current analysis is the first step of a broad physics programme on neutrino measurements at the LHC, which will provide important insights in neutrino and electroweak physics, as well as in quantum chromodynamics by probing forward hadron production and the deep inelastic scattering of high-energy neutrinos, as detailed in Refs.~\cite{Anchordoqui:2021ghd, Feng:2022inv,Cruz-Martinez:2023sdv, Kling:2023tgr, Wilkinson:2023vvu}.

\section{The FASER$\nu$ detector and data taking}

The FASER$\nu$ emulsion detector is made of 730 layers of interleaved tungsten plates and emulsion films~\cite{Ariga:2020lbq}, with a total target mass of 1.1~tonnes. 
The tungsten plates are 1.1~mm thick, and each emulsion film is 0.34~mm thick. The detector is 1.05~m long and has a transverse area with respect to the neutrino beam of 25 $\times$ 30~cm$^2$. 
The detector is aligned with the LOS and placed in front of the FASER spectrometer, 480~m away from the $pp$ collision point within the ATLAS experiment (IP1). 
A more detailed description of the FASER$\nu$ detector is provided in Ref.~\cite{FASER:2022hcn}.

\begin{figure}[h]
\centering
\includegraphics[width=0.95\linewidth]{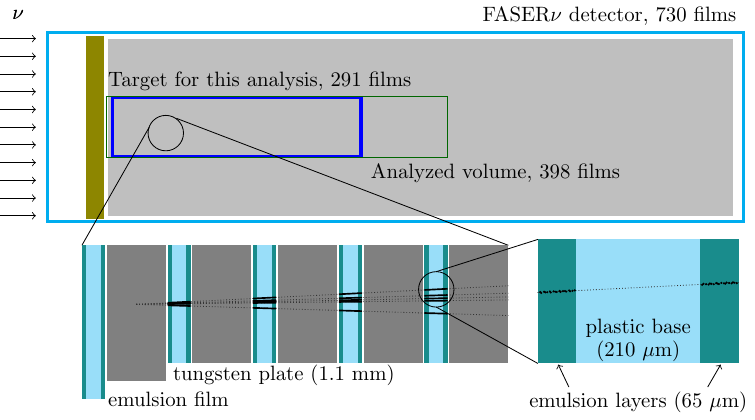}
\caption{Schematic view of the analysed detector volume (side view). The FASER$\nu$ box contains a total of 730 emulsion films and is shown in grey. The thin green box outlines the reconstructed volume, and neutrino interactions are searched for within the fiducial volume defined by the blue box.}
\label{fig:analysed_volume}
\end{figure}

The analysed dataset was collected between July 26 and September 13, 2022, corresponding to 9.5~fb$^{-1}$ of $pp$ collisions at a centre-of-mass energy of 13.6~TeV. The integrated luminosity is measured by the ATLAS experiment~\cite{ATL-DAPR-PUB-2023-001,ATLAS:2022hro,Avoni:2018iuv} with an uncertainty of 2.2\%.
For the analysis presented in this paper, only 14\% of the detector volume, shown in Figure~\ref{fig:analysed_volume}, is considered as the target region for neutrino interaction vertices. In the transverse\footnote{A Cartesian coordinate system is used with the $z$-axis running along the LOS from the ATLAS collision point to FASER, the $y$-axis pointing vertically upwards, and the $x$-axis pointing horizontally to the centre of the LHC ring. The angle $\phi$ is the azimuthal angle in the ($x$-$y$) plane, and $\theta$ is the polar angle measured from the beam direction.} 
($x$-$y$) plane, a region of 23.4~cm $\times$ 9.0~cm is analysed, and in the longitudinal direction, 41.5~cm, including 31.6~cm of tungsten (291 tungsten plates), is considered. The corresponding target mass is 128.6~kg.
Data from seven films upstream of the target region are used to check the absence of charged parent tracks. 
Data from an additional 100 plates immediately downstream of the target region are used to measure the energy or the momentum of the particle tracks. 
The LOS passes through the centre of the analysed volume in the horizontal plane and about 2~cm from the bottom in the vertical plane. 

After chemical development, the emulsion films are read out by the Hyper Track Selector (HTS) system~\cite{Yoshimoto:2017ufm}. 
The HTS system divides each emulsion film into eight readout zones. 
A multi-stage alignment procedure is used. The first alignment between each two consecutive films is performed using the recorded track hits for each readout zone. The data are then divided into sub-volumes of 1.7~cm $\times$ 1.7~cm $\times$ 15 films due to limitations in memory usage, and to achieve a better alignment resolution. After a second alignment between each two consecutive films and track reconstruction in each sub-volume, an additional alignment is applied by selecting those tracks crossing 15 plates to improve the tracking resolution. Finally, the track reconstruction procedure~\cite{DsTau:2019wjb} links track hits on different films by correlating their positions and angles.

\section{Simulation samples}

Monte Carlo (MC) samples for neutrino and background processes are used to define the event selection criteria, to estimate the backgrounds, and to assess systematic uncertainties. 

The neutrino energy spectra, relative flavour composition, and transverse spatial distribution are simulated as follows. 
The neutrino fluxes are obtained using the fast neutrino flux simulation introduced in Ref.~\cite{Kling:2021gos}. As described in Ref.~\cite{FASER:2024ykc}, light hadrons (pions, kaons, and hyperons) are simulated using \texttt{EPOS-LHC}~\cite{Pierog:2013ria}, \texttt{QGSJET~II-04}~\cite{Ostapchenko:2010vb}, \texttt{SIBYLL2.3d}, and \texttt{PYTHIA8}~\cite{Bierlich:2022pfr} with a forward physics tune~\cite{Fieg:2023kld}. \texttt{EPOS-LHC} is used as the baseline flux model, with the envelope of the considered models used to determine the systematic uncertainty. Charm hadrons are simulated using \texttt{POWHEG+PYTHIA8}~\cite{Buonocore:2023kna}. 
In Ref.~\cite{Buonocore:2023kna}, the uncertainty is considered by varying the renormalization and factorization scale by a factor of two. 
The interaction of neutrinos with the tungsten/emulsion detector is simulated using the \texttt{GENIE} event generator~\cite{Andreopoulos:2009rq,Andreopoulos:2015wxa}. The interactions of all other particles traversing the tungsten/emulsion detector are simulated using \texttt{GEANT4}~\cite{GEANT4:2002zbu}. In the default setup, the \texttt{FTFP\_BERT} physics list is used to model hadron interactions.

The main background contributions arises from neutral hadrons produced in the photo-nuclear interactions of muons within the rock in front of the FASER detector or within the FASER$\nu$ detector material. 
The muon flux is measured to be $(1.43 \pm 0.07) \times 10^4$ tracks/cm$^2$/fb$^{-1}$, as estimated from the reconstructed tracks within an angle of $\Delta \theta<$10 mrad from the beam direction in the FASER$\nu$ detector. 
The neutral-hadron background is simulated through a multi-step process. First, the energy spectrum of muons is simulated using the \texttt{FLUKA} package~\cite{Ferrari:2005zk,Battistoni:2015epi}, which includes a detailed model of the LHC infrastructure between IP1 and FASER. \texttt{GEANT4} was then used to simulate the interactions of these muons within the rock in front of FASER or in the tungsten of the FASER$\nu$ detector. Finally, high-statistics samples of the individual neutral-hadron species ($K_S$, $K_L$, $n$, $\Lambda$, $\bar{n}$, and $\bar{\Lambda}$) were produced and weighted to follow the expected energy spectra estimated by the previous step.

The simulated events are reconstructed in the same way as the data. Corrections are applied to the simulated samples to reproduce the hit efficiencies, position and angular resolutions of the data.

\section{Event reconstruction and selection}

Candidate $\nu_e$ and $\nu_{\mu}$ CC interactions are selected based on reconstructed charged particle tracks, forming a neutral vertex in FASER$\nu$. 
The number of neutral hadrons drops quickly with increasing energy. 
Since neutrinos are more energetic than the neutral-hadron background, the tracks associated to the vertex are boosted in the forward direction. A CC interaction produces a high-energy electron or muon, well separated in azimuthal angle from the other particles associated with the vertex.

\subsection{Vertex selection}

Using the reconstructed tracks passing through at least three plates, the vertex reconstruction is performed by searching for converging track patterns with an impact parameter less than 5~$\mu$m. Tracks with $\tan \theta \leq 0.5$ are retained, and converging patterns with more than four tracks are selected as vertices. The tracks are required to start within three films downstream of the vertices. The number of tracks with $\tan \theta \leq 0.1$ relative to the beam direction is also required to be greater than three to suppress the neutral-hadron background. Furthermore, the vertices are required to not have a parent track.

\subsection{Electron identification and energy measurement}

Candidate $\nu_e$ CC interactions are selected from the initial set of vertices by requesting an associated high-energy electromagnetic (EM) shower with a reconstructed energy above 200~GeV and $\tan \theta >$ 0.005. The latter requirement is used to reduce the neutral-hadron background. It improves the signal-to-noise ratio because leptons from neutrino CC interactions have larger angles, due to higher transverse momenta, than high-energy particles from the neutral hadron background.
The EM shower is required to start within two films downstream of the vertices.
The EM shower must have an azimuthal angle, relative to the sum of all other tracks in the vertex, $\Delta \phi > \pi/2$. This cut is motivated by the expectation that, for CC interactions, the outgoing lepton and hadrons are back to back in the transverse plane.

The EM shower is formed by reconstructed tracks in the emulsion, produced from electron/positron pair production as the shower develops~\cite{Kobayashi:2012jb, Juget:2009zz}. Given the short radiation length of tungsten, the EM shower remains compact and can be associated with the vertex. 
The EM shower reconstruction is based on searching for track segments in a cylinder of radius 100 $\mu$m around the shower axis, defined by iteratively fitting the group of segments belonging to the shower. 
The algorithm selections were defined using the profile of EM showers from electrons with energy above 200 GeV in simulation and that of background showers in data and simulation.
The number of track segments in $\pm 3$ films around the shower maximum (seven films in total), which we denote $N_{\text{seg}}$, is used to identify EM showers and estimate the electron energy. 
Several selections on the track segment position and angle with respect to the shower axis are applied to reduce the contribution from background segments not associated with the EM shower.
Additionally, the average residual background is estimated by using randomly positioned cylinders, whose average number of segments, $N_{\text{seg}}^{\text{bg}}$ is then subtracted from the shower before the energy is estimated. 
The energy reconstruction is performed using the relation between the difference $N_{\text{seg}}- N_{\text{seg}}^{\text{bg}}$ and electron energy which is fitted with a polynomial, as defined with simulation.
The energy reconstruction algorithm performance was tested for electrons in the $\nu_e$ MC simulation, showing a resolution of around 25\% at 200~GeV and between 25-40\% at higher energies.

\subsection{Muon identification and momentum measurement}

Candidate $\nu_{\mu}$ CC interactions are selected from the initial set of vertices by requiring that one of the reconstructed charged particle tracks associated with the vertex is a muon candidate. 
Muon candidates are defined as tracks that penetrate more than 100 tungsten plates without exhibiting secondary hadron interactions consisting of at least two daughter tracks. Furthermore, the muon reconstructed momentum $p$ has to be greater than 200~GeV with $\tan \theta >$ 0.005. Simulation studies show that the probability for a charged hadron with $p>$ 200 GeV to satisfy the muon candidate requirements is about 20\%. The muon candidate track is then required to have an azimuthal angle $\Delta \phi > \pi/2$.

The track momentum is estimated by measuring multiple Coulomb scattering using the so-called coordinate method~\cite{Kodama:2007mw}. The performance of the momentum evaluation algorithm is studied using simulated muon tracks with a flat momentum distribution from 1 to 2000 GeV, including position and angular smearing to account for residual misalignments between the emulsion films. The resolution, as quantified by the RMS of the distribution of the difference between the true momentum and reconstructed momentum, is around 30\% at 200~GeV and reaches 50\% at higher energies.

\subsection{Selection efficiencies and systematic uncertainties}

\begin{figure}[h]
\centering
\includegraphics[width=1.00\linewidth]{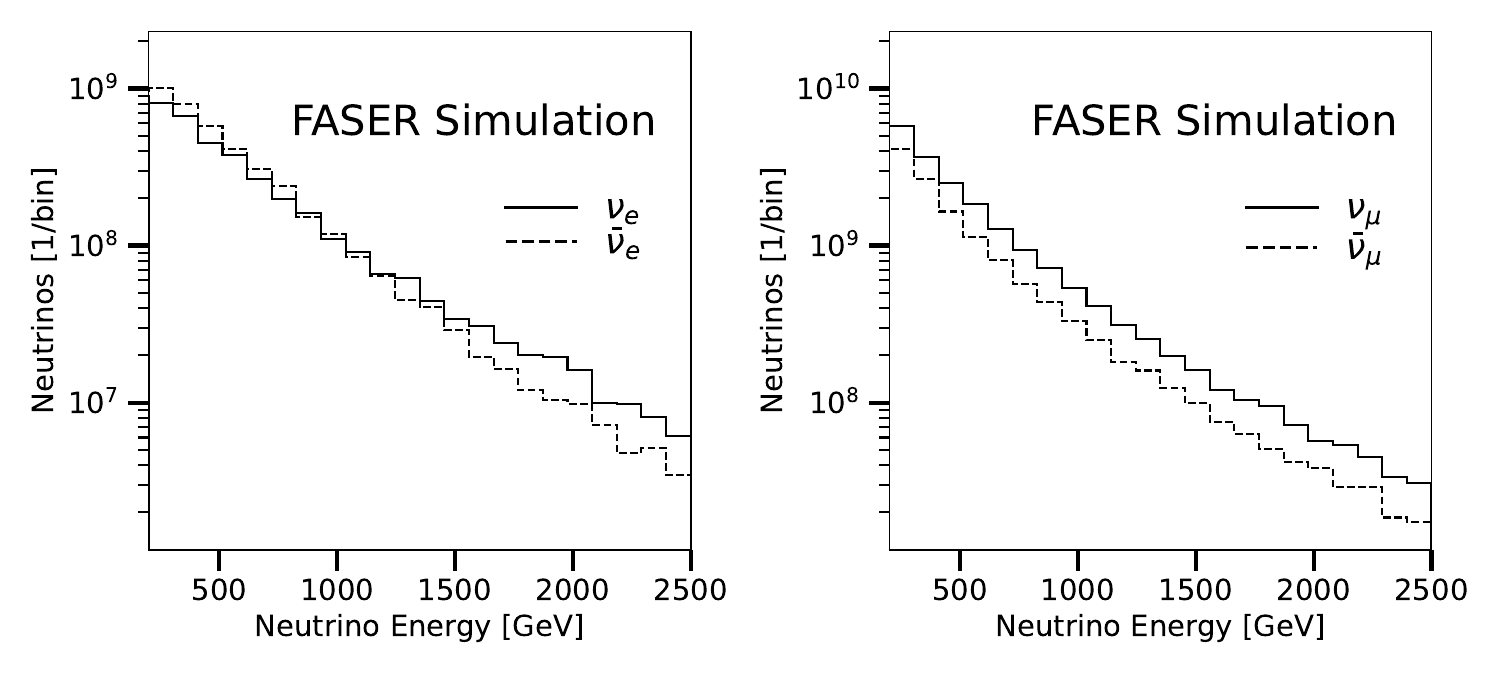}
\includegraphics[width=0.95\linewidth]{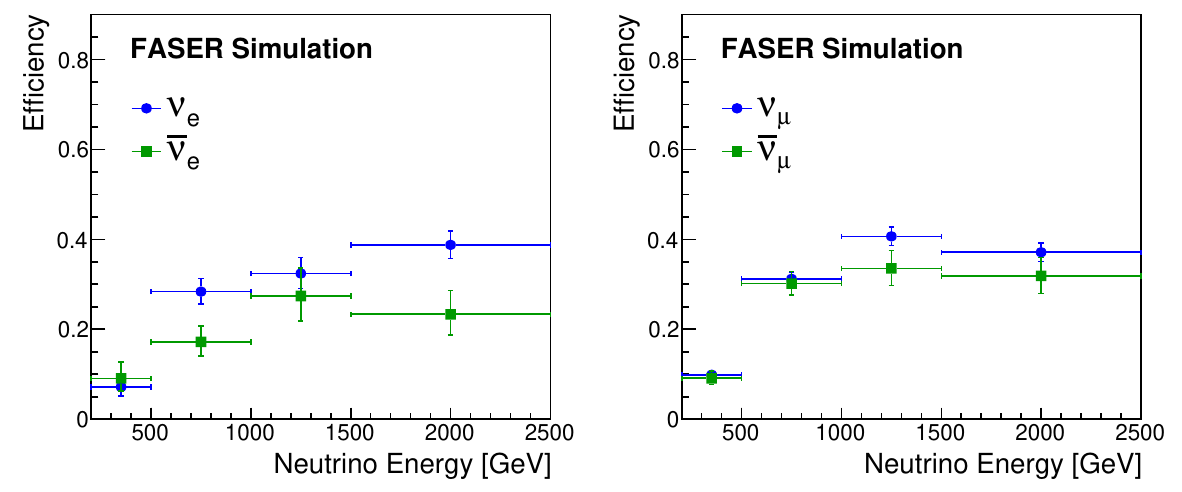}
\caption{Top: Simulation of the fluxes for $\nu_e$ and $\bar{\nu}_e$ (left), and for $\nu_\mu$ and $\bar{\nu}_\mu$ (right). Bottom: Selection efficiencies for $\nu_e$ CC and $\bar{\nu}_e$ CC interactions (left), and for $\nu_\mu$ CC and $\bar{\nu}_\mu$ CC interactions (right). The statistical uncertainties are shown.
}
\label{fig:selection_efficiencies}
\end{figure}

The selection efficiencies for $\nu_e$ CC, $\bar{\nu}_e$ CC, $\nu_\mu$ CC, and $\bar{\nu}_\mu$ CC events are shown in Figure~\ref{fig:selection_efficiencies}. 
Due to the helicity combination, leptons in anti-neutrino events are more boosted, and the other particles have less energy, than in neutrino events. Consequently, the efficiency of anti-neutrino events to pass the vertex selection is slightly lower than that of neutrino events. 

Systematic uncertainties related to the signal expectation are summarized in Table~\ref{tab:systematic_uncertainty_n_exp}. 
The systematic uncertainty from tungsten thickness is estimated based on the variation of the measured plate thickness. The systematic uncertainty from the LOS position is estimated by varying the detector alignment with the LOS by $\pm$1~cm. 
The systematic uncertainty from hadronization in the neutrino interaction simulation is estimated using five different \texttt{PYTHIA} physics tunes, varying the hadronization parameters~\cite{GENIE:2021wox}. 
The systematic uncertainty related to the reconstruction, including kinematical measurements, is estimated by varying the track segment efficiency from the nominal value of 90\% to the worst case of 80\%. Other systematic uncertainties in the reconstruction are sub-dominant, and an overall reconstruction systematic uncertainty of 20\% is assigned.

The systematic uncertainties listed in Table~\ref{tab:systematic_uncertainty_n_exp} for $\nu_e$ are dominated by the flux uncertainty, with the hadronization and reconstruction uncertainties contributing at the 20\% level. The flux uncertainty is dominant for $\nu_e$ since a significant fraction of $\nu_e$ originates from decays of charm hadrons, which have large uncertainties in their forward production. For $\nu_\mu$, the flux uncertainty is sub-dominant, with the hadronization and reconstruction uncertainties dominating.

\renewcommand{\arraystretch}{1.5}
\begin{table}[h]
\centering
\caption{Systematic uncertainties related to the signal expectation.}
\begin{center}
\small
{
\begin{tabular}{lll}
\hline
Source & \multicolumn{2}{l}{Relative uncertainty} \\
\      & \normalsize $\nu_e$ & \normalsize $\nu_\mu$ \\
\hline
Luminosity                    & {\footnotesize 2.2\%}  & {\footnotesize 2.2\%}  \\
Tungsten thickness            & {\footnotesize 1\%}    & {\footnotesize 1\%}    \\
Interactions with emulsions   & {$^{+3.6}_{-0}$\%}     & {$^{+3.6}_{-0}$\%}     \\
\hline
Flux uncertainty              & {$^{+70}_{-22}$\%}     & {$^{+16}_{-9}$\%}      \\
Line of sight position        & {$^{+2.1}_{-2.4}$\%}   & {$^{+1.9}_{-2.5}$\%}   \\
Efficiency from hadronization & {$^{+22}_{-5}$\%}      & {$^{+23}_{-5}$\%}      \\
Efficiency from reconstruction& {\footnotesize 20\%}   & {\footnotesize 20\%}   \\
Efficiency from MC statistics & {\footnotesize 4.9\%}  & {\footnotesize 2.8\%}  \\
\hline
Total                         & $^{+70}_{-22}$\% \footnotesize(flux)  & $^{+16}_{-9}$\% \footnotesize(flux) \\
\                             & $^{+30}_{-21}$\% \footnotesize(other) & $^{+31}_{-21}$\% \footnotesize(other) \\
\hline 
\end{tabular}
}
\label{tab:systematic_uncertainty_n_exp}
\end{center}
\end{table}
\renewcommand{\arraystretch}{1}

\section{Backgrounds}
\label{sec:backgrounds}

The background from neutral-hadron interactions is estimated using MC simulation. 
The neutral-hadron MC samples are normalized to the equivalent luminosity of the data by using the number of observed and simulated muons. The final neutral-hadron samples are equivalent to $\sim$400 times the size of the data.

Table~\ref{tab:neutral_hadron_summary} shows the selected number of neutral-hadron simulated events when applying the $\nu_{e}$ and $\nu_{\mu}$ CC selections.

\begin{table}[h]
    \centering
    \caption{The number of MC reconstructed events of neutral-hadron interactions satisfying the $\nu_e$ and $\nu_\mu$ CC event selection. The scaling factor shows the ratio of the data luminosity to the MC luminosity.
    }
    \small
    \begin{tabular}{lcccccc}
         \\
         Hadron type & $K_L$ & $n$ & $\Lambda$\\
         \hline 
         Events simulated ($E_h>$200 GeV) & 13497 & 13191 & 13902 \\
         Events selected as $\nu_e$ CC & 0 & 0 & 0 \\
         Events selected as $\nu_\mu$ CC & 6 & 11 &5 \\
         Scaling factor (data/MC) & 1/232 & 1/256 & 1/423 \\
         \\
         Hadron type & $K_S$ & $\bar{n}$ & $\bar{\Lambda}$\\
         \hline 
         Events reconstructed ($E_h>$200 GeV) & 7113 & 5827 & 5368 \\
         Events selected as $\nu_e$ CC & 1 & 0 &0 \\
         Events selected as $\nu_\mu$ CC & 3 & 3 & 4 \\
         Scaling factor (data/MC) & 1/436 & 1/569 & 1/630 \\
         \hline          
    \end{tabular}
    \label{tab:neutral_hadron_summary}
\end{table}

\begin{figure*}[tb]
\centering
\includegraphics[width=0.90\linewidth]{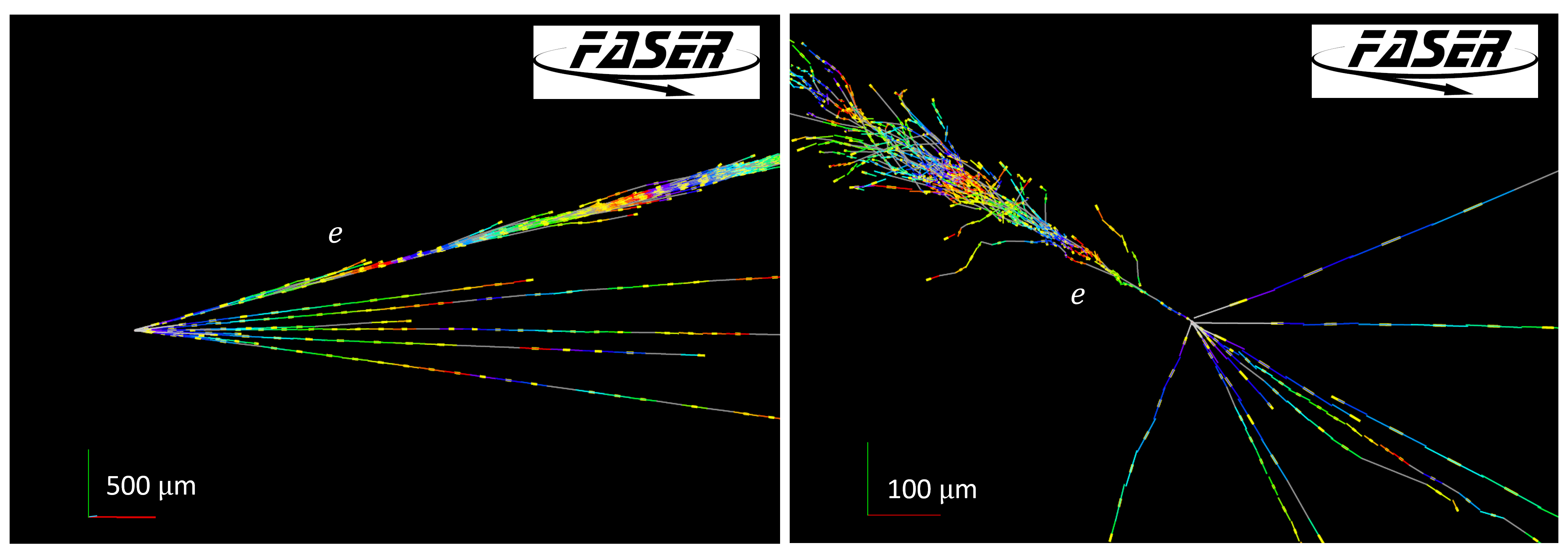}
\includegraphics[width=0.90\linewidth]{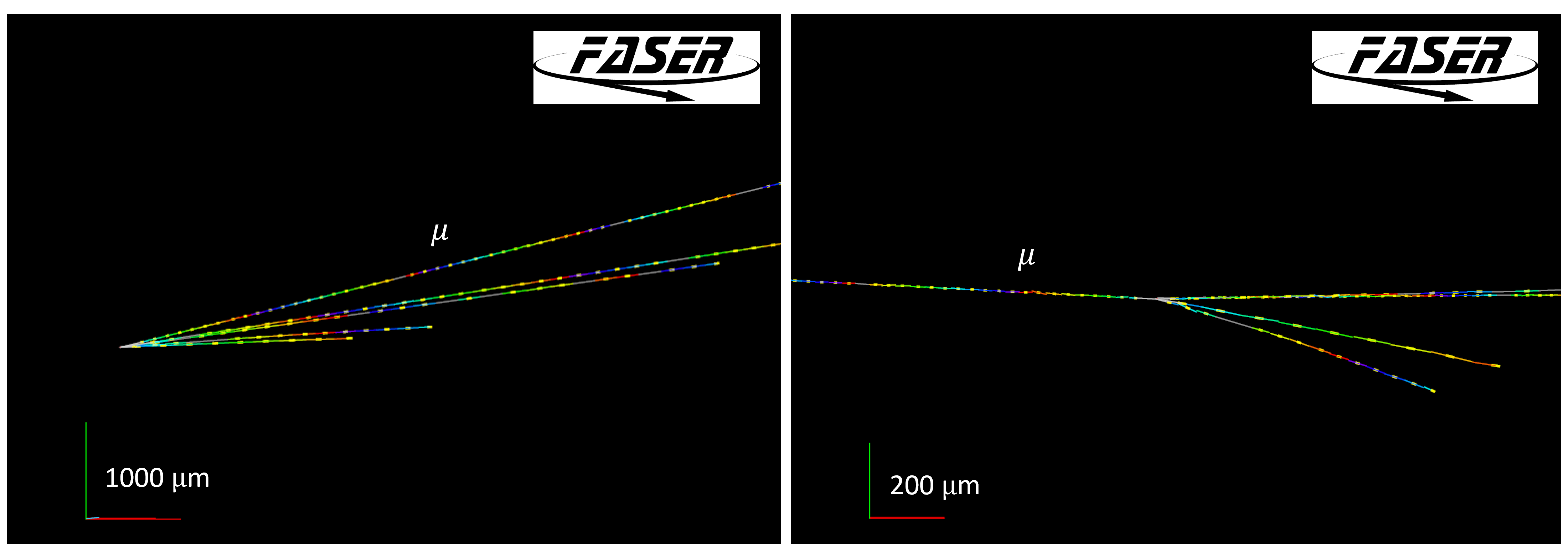}
\caption{Event displays of one of the $\nu_e$ CC candidate events (top) and one of the $\nu_{\mu}$ CC candidate events (bottom). 
In each panel, the right-handed coordinate axes are shown in the bottom left, with red, green, and blue axes indicating the $x$ (horizontal), $y$ (vertical), and $z$ (beam) directions, respectively.  The right panels show views transverse to the beam direction, and so the blue axes are not visible. The left panels are slightly rotated views, with the blue axes barely visible, to show the longitudinal development of the event. Yellow line segments show the trajectories of charged particles in the emulsion films. The other coloured lines are interpolations, with the colours indicating the longitudinal depth in the detector.
}
\label{fig:event_displays}
\end{figure*}

Systematic uncertainties on the neutral-hadron background estimate are evaluated by varying the incident muon energy distribution and by varying the physics lists used to model the neutral-hadron interactions in \texttt{GEANT4}. The incoming muon energy distribution was scaled up and down by a factor of $1+E/(3~\text{TeV})$ to distort the spectrum as a function of muon energy $E$, and the effect on the expected neutral-hadron background was evaluated. 
In addition, the relative change in the background was checked using the physics list \texttt{QGSP\_BERT}~\cite{Allison:2016lfl} to model the hadron interactions instead of the \texttt{FTFP\_BERT} physics list. 
From these studies, a systematic uncertainty of 100\% on the expected background is assigned.

\begin{spacing}{1.05}
In addition to the neutral-hadron background, there is a contribution to the set of vertices retained by the $\nu_e$ and $\nu_{\mu}$ CC selection from NC neutrino interactions. 
The background from NC neutrino interactions is estimated from simulated samples. None of the simulated NC events passed the $\nu_e$ CC selection, using a sample equivalent to 300 times the size of the analyzed dataset. 
The number of NC events expected in the analysed dataset after the $\nu_{\mu}$ CC selection is estimated as \NncContamiLight\ and \NncContamiCharm\ for events originating from light hadrons and charm hadrons, respectively. 

The total background estimates are \BackgroundRandomNuE\ and \BackgroundRandomNuMu\ for the $\nu_e$ and $\nu_\mu$ selections, respectively.
\end{spacing}

\section{$\nu_e$ and $\nu_{\mu}$ candidate events}

Four events are selected by the $\nu_{e}$ selection on data. The highest reconstructed electron energy from the selected $\nu_e$ CC candidates is 1.5~TeV. It is therefore the highest-energy $\nu_e$ interaction ever detected by accelerator-based experiments. 

Eight events are selected by the $\nu_{\mu}$ selection on data. The highest reconstructed muon momentum from the selected $\nu_{\mu}$ CC candidates is 864~GeV, meaning that the $\nu_{\mu}$ sample includes neutrinos with energy likely above 1 TeV, far higher than from previous accelerator-based neutrino studies.

Example event displays of $\nu_e$ and $\nu_\mu$ candidates are shown in Figure~\ref{fig:event_displays}. 
As expected, both events exhibit a back-to-back topology between the lepton candidate and the other tracks in the vertex. 

The expected number of neutrino signal events satisfying the selections are in the range 1.1--3.3 (for $\nu_e$ CC) and 6.5--12.4 (for $\nu_{\mu}$ CC), where the range covers the uncertainties listed in Table~\ref{tab:systematic_uncertainty_n_exp}. The observed number of interactions is consistent with Standard Model predictions.

\begin{figure*}[htb]
\centering
\subfigure{
\includegraphics[height=5.9cm]{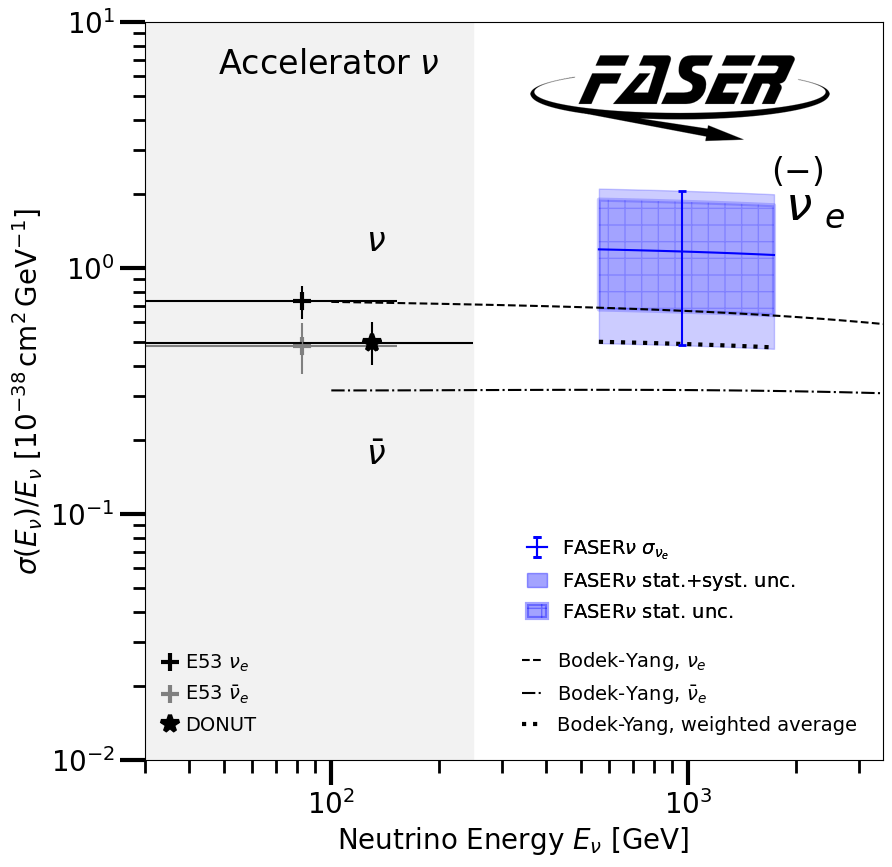}
}
\subfigure{
\includegraphics[height=5.9cm]{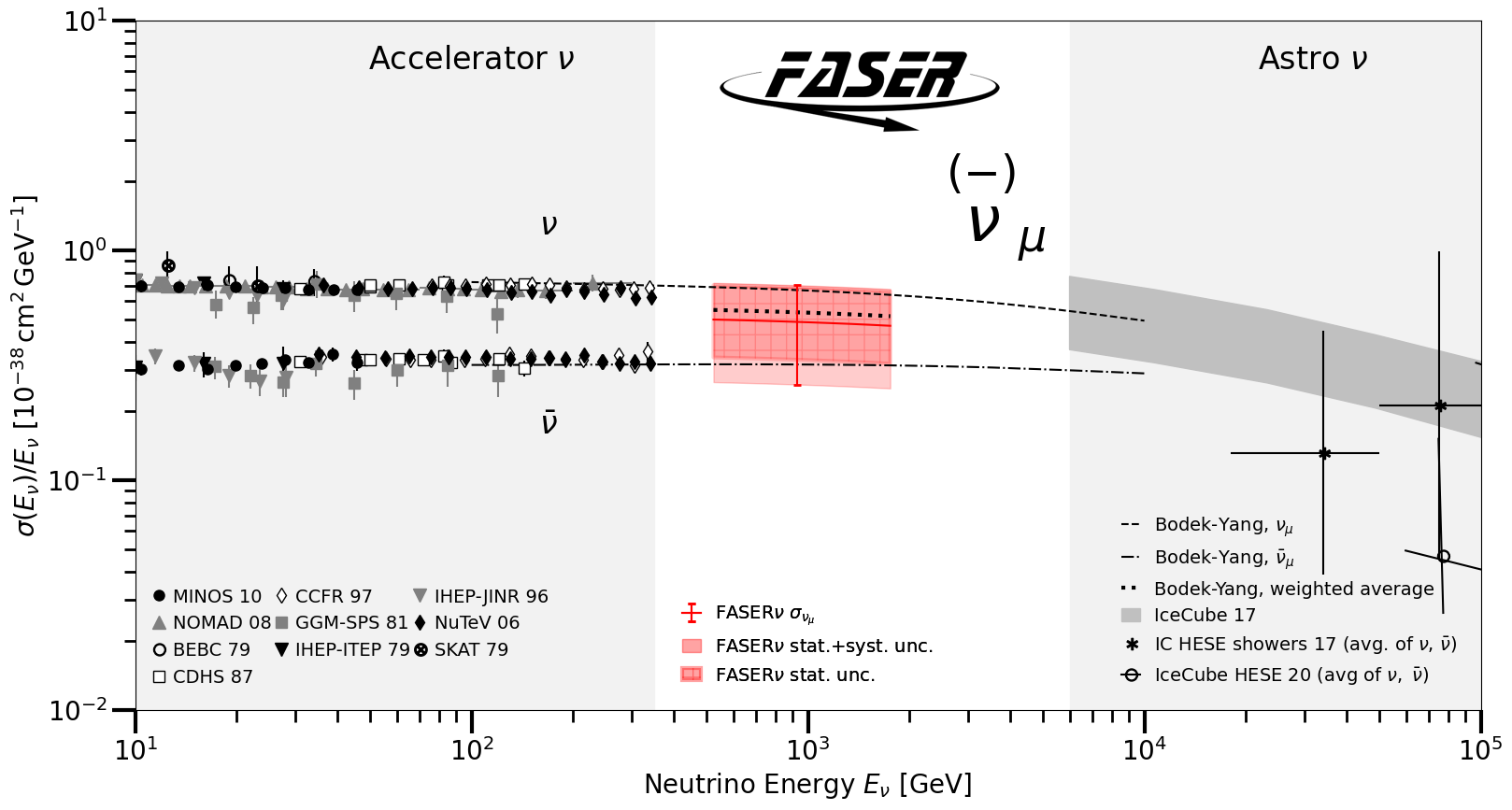}
}
\caption{The measured cross section per nucleon for $\nu_e$ (left) and $\nu_{\mu}$ (right). The dashed contours labelled ``Bodek-Yang'' are cross sections predicted by the Bodek-Yang model, as implemented in \texttt{GENIE}. Note that the displayed experiments do not all use the same targets.
}
\label{fig:measured_cross_sections}
\end{figure*}

\begin{spacing}{1.1}
The statistical significance of the observation of $\nu_e$ and $\nu_{\mu}$ is estimated by considering the confidence level for excluding the null hypothesis (background-only). 
Based on the Probability Density Function (PDF) for the background, 10$^{10}$ pseudo-experiments were generated. 
The neutral-hadron background was generated following separate Poisson distributions for each neutral hadron species considered, with a Gaussian-distributed systematic uncertainty of 100\% included.
The background from neutrino NC events was generated from a Poisson distribution, with systematic uncertainties included via Gaussian-distributed nuisance parameters (separately for the uncertainties from the light hadron ($\lambda_{\rm light\,hadron}$) and charm hadron ($\lambda_{\rm charm\,hadron}$) neutrino flux, and the experimental uncertainties ($\lambda_{\rm syst}$)).

A random value, $N$, was calculated following the total background PDF, and the number of pseudo-experiments with $N \geq N_{\rm obs}$ was counted, where $N_{\rm obs}$ is the number of observed neutrino events. Based on the fraction of cases with $N \geq N_{\rm obs}$, 
observed p-values of $8.8\times 10^{-8}$ for $\nu_e$ and $5.7\times 10^{-9}$ are obtained, corresponding to significances of \SignificanceNuEFullToy~$\sigma$ for $\nu_e$ and \SignificanceNuMuFullToy~$\sigma$ for $\nu_{\mu}$ for the exclusion of the null hypothesis. 
The expected significance is estimated with pseudo-experiments with the signal expectation from the baseline flux model to be 3.3~$\sigma$ for $\nu_e$ and 6.4~$\sigma$ for $\nu_{\mu}$.
\end{spacing}

\section{$\nu_e$ and $\nu_{\mu}$ cross sections}

The number of observed neutrino events can be described as 
\begin{align*}
& N_{\rm obs}=\frac{L \, \rho \, l}{m_{\rm nucleon}}\int \sigma(E) \, \phi(E) \, \varepsilon(E) \, \mathrm{d} A \, \mathrm{d}E, 
\end{align*}
where $L$ is the luminosity, $\rho$ is the density of tungsten (19.3 g/cm$^3$), $l$ is the thickness of the tungsten plates, $m_{\rm nucleon}$ is the mass of the nucleon, $\sigma(E)$ is the cross section, $\phi(E)$ is the neutrino flux at the detector integrated over the transverse area $A$ and the energy $E$, and $\varepsilon(E)$ is the detection efficiency.

The $\nu_e$ and $\nu_\mu$ CC cross sections are measured in a single energy bin. The ratio between the cross sections evaluated with the \texttt{GENIE} simulation ($\sigma_{\rm theory}$) and with the observed data is defined as a factor $\alpha$ as described by $\sigma_\textnormal{obs} = \alpha \cdot \sigma_\textnormal{theory}$, assuming that $\alpha$ is common for neutrino and anti-neutrino interactions. 
The energy range for $\sigma_{\rm theory}$ was defined to contain 68\% of reconstructed neutrinos using the baseline models, which is 560--1740 GeV and 520--1760 GeV for $\nu_e$ and $\nu_\mu$, respectively. 

The PDF of $\alpha$ from a Bayesian method~\cite{ParticleDataGroup:2020ssz} can be calculated by integrating the product of the likelihood, $\mathcal L$, and the prior probability distribution of the nuisance parameters, $\pi$. 
The likelihood, $\mathcal L$, is constructed using the parameters $N_{\rm obs}$, $\alpha$, $\lambda_{\rm light\,hadron}$, $\lambda_{\rm charm\,hadron}$, and $\lambda_{\rm syst}$ that were defined in the previous section, the nuisance parameters for each neutral-hadron background and its systematic uncertainty, and the nuisance parameters for the NC background. The prior distribution, $\pi$, is flat and assumed to be 1 except for unphysical values of the parameters where it is 0.
The posterior is obtained with a Markov chain MC with the Metropolis-Hastings algorithm~\cite{gregory_2005}. 

The $\alpha$ parameter is measured to be $\AlphaResultNuE$ ($\AlphaResultNuMu$) for $\nu_e$ ($\nu_\mu$), respectively, where statistical and systematic uncertainties are combined. The statistical uncertainty is $_{-1.0}^{+1.4}$ ($_{-0.3}^{+0.4}$), which dominates the uncertainty of the measurement. The next most important component is the flux uncertainty, estimated in a fit where only the flux nuisance parameter is free and equal to $_{-0.8}^{+0.4}$ ($_{-0.08}^{+0.03}$), after quadratically subtracting the statistical uncertainty. The other systematic uncertainties are $_{-0.7}^{+0.9}$ ($_{-0.2}^{+0.3}$). The sum of these individual uncertainties should include correlations to result in the total uncertainty. 
The energy-independent part of the interaction cross sections per nucleon, $\sigma_{\rm obs}/E_{\nu}$, is measured over the considered energy ranges to be ($\XsecObsResultNuE$) $\times 10^{-38}$ $\mathrm{cm}^{2}\,\mathrm{GeV}^{-1}$ for $\nu_e$ and ($\XsecObsResultNuMu$) $\times 10^{-38}$ $\mathrm{cm}^{2}\,\mathrm{GeV}^{-1}$ for $\nu_\mu$. 
Figure~\ref{fig:measured_cross_sections} shows the measured cross sections, together with those obtained by other experiments~\cite{Baltay:1988au, DONuT:2007bsg, MINOS:2009ugl, NOMAD:2007krq, T2K:2013nor, T2K:2014hih, T2K:2014axs, ArgoNeuT:2011bms, ArgoNeuT:2014rlj, Barish:1978pj, deGroot:1978feq, Colley:1979rt, Baker:1982ty, Berge:1987zw, Seligman:1997fe, GargamelleNeutrinoPropane:1979kqo, GargamelleSPS:1981hpd, Mukhin:1979bd, Anikeev:1995dj, NuTeV:2005wsg, SciBooNE:2010slc, Baranov:1978sx, IceCube:2017roe, Bustamante:2017xuy, IceCube:2020rnc}.
The measured value of $\sigma_{\rm obs}$ is shown as the blue curved line for $\nu_e$ and the red curved line for $\nu_{\mu}$. 
The weighted average of the \texttt{GENIE}-predicted cross section is also shown, assuming the ratio of the incoming neutrino to anti-neutrino fluxes to be 1.04 for $\nu_e$ and 0.61 for $\nu_{\mu}$.

\section{Conclusions}

First results from the search for high-energy electron and muon neutrino interactions in the FASER$\nu$ tungsten/emulsion detector of the FASER experiment have been presented. 
The analysis uses a subset of the FASER$\nu$ volume, corresponding to a target mass of 128.6~kg, exposed to 9.5~fb$^{-1}$ of LHC $pp$ collisions during the summer of 2022. 
Selections are applied to retain reconstructed vertices consistent with high-energy $\nu_e$ and $\nu_\mu$ CC interactions, while minimizing the background from neutral-hadron interactions. 
Four electron neutrino interaction candidate events are observed, with an expected background of \BackgroundRandomNuE, leading to a statistical significance of \SignificanceNuEFullToy~standard deviations. This represents the first direct observation of electron neutrinos produced at a particle collider. 
Eight muon neutrino interaction candidate events are also found, with an expected background of \BackgroundRandomNuMu, leading to a statistical significance of \SignificanceNuMuFullToy~standard deviations. 
The interaction cross section per nucleon is measured over an unexplored energy range of 560--1740 GeV for $\nu_e$ and 520--1760 GeV for $\nu_{\mu}$. In these energy ranges, the neutrino-antineutrino combined cross sections, $\sigma_{\rm obs}/E_{\nu}$, are constrained to be ($\XsecObsResultNuE$) $\times 10^{-38}$ $\mathrm{cm}^{2}\,\mathrm{GeV}^{-1}$ for $\nu_e$ and ($\XsecObsResultNuMu$) $\times 10^{-38}$ $\mathrm{cm}^{2}\,\mathrm{GeV}^{-1}$ for $\nu_{\mu}$, both consistent with the cross section predictions of the Standard Model. 
These results demonstrate the capability to study flavour-tagged neutrino interactions at TeV energies with the FASER$\nu$ emulsion-based detector at the LHC.

\section*{Acknowledgements}

We thank CERN for the excellent performance of the LHC and the technical and administrative staff members at all FASER institutions for their contributions to the success of the FASER experiment. We thank the ATLAS Collaboration for providing us with accurate luminosity estimates for the Run 3 LHC $pp$ collision data. We also thank the CERN STI group for providing detailed \texttt{FLUKA} simulations of the muon fluence along the LOS, the CERN EN-HE group for their help with the installation/removal of the FASER$\nu$ detector, and the CERN EP department for the refurbishment of the CERN dark room, used for the preparation and development of the FASER$\nu$ emulsion. 
In addition, we thank Nicolas Berger for reviewing the statistical analysis used, Mauricio Bustamante for providing the digitized neutrino cross section data, and Saya Yamamoto for supporting the preparation of emulsion films.

This work was supported in part by Heising-Simons Foundation Grant Nos.~2018-1135, 2019-1179, and 2020-1840, Simons Foundation Grant No.~623683, U.S. National Science Foundation Grant Nos.~PHY-2111427, PHY-2110929, and PHY-2110648, JSPS KAKENHI Grant Nos.~19H01909, 22H01233, 20K23373, 23H00103, 20H01919, and 21H00082, the joint research program of the Institute of Materials and Systems for Sustainability, ERC Consolidator Grant No.~101002690, BMBF Grant No.~05H20PDRC1, DFG EXC 2121 Quantum Universe Grant No.~390833306, Royal Society Grant No.~URF$\backslash$R1$\backslash$201519, UK Science and Technology Funding Councils Grant No.~ST/ T505870/1, the National Natural Science Foundation of China, Tsinghua University Initiative Scientific Research Program, and the Swiss National Science Foundation.

\clearpage
\appendix

\section{Appendix A: Reconstructed data quality}
\begin{figure}[htb]
\centering
\includegraphics[width=\linewidth]{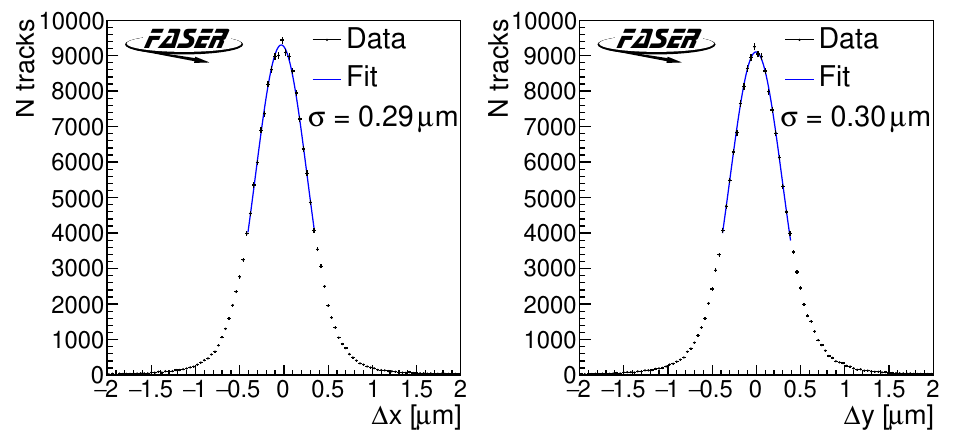}
\vspace*{-3mm}
\caption{Distributions of the position deviation of the track hits with respect to a linear-fit line for the reconstructed tracks, measured in a typical sub-volume of the FASER$\nu$ data. The blue line shows a Gaussian fit to the data, giving a position resolution of 0.29--0.30~$\mu$m.
}
\label{fig:position_resolution}
\end{figure}

Appendix A describes the reconstructed data quality of the analyzed volume, checked using penetrating muon tracks. Figure~\ref{fig:position_resolution} shows the measured position resolution of 0.3 $\mu$m in a plane transverse to the beam. 
The track hit efficiency in each film was measured to be greater than 90\% in the reconstructed volumes, corresponding to an efficiency greater than 99.98\% for detecting tracks with at least three hits in seven films. 

\vspace{-2mm}
\section{Appendix B: Expected energy spectrum of interacting neutrinos}

\begin{figure}[hbt]
\centering
\includegraphics[width=0.87\linewidth]{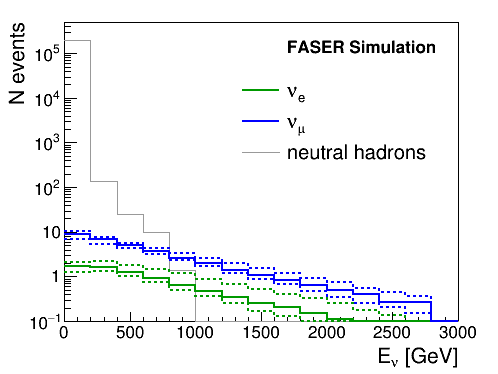}
\vspace*{-3mm}
\caption{
Simulation of the energy spectrum of electron (solid green curve) and muon neutrinos (solid blue curve) within the analysed detector volume are shown for an integrated luminosity of 9.5 fb$^{-1}$. The dashed lines envelope the predictions from different generators; for more details, see text. The energy spectrum of neutral hadrons (grey curve) is also shown.
}
\label{fig:E_hadron_and_nu}
\end{figure}

Appendix B describes the expected energy spectrum of interacting neutrinos, as well as that of the neutral-hadron backgrounds; these are shown in Figure~\ref{fig:E_hadron_and_nu}. The expected numbers of neutrino interaction events before selection cuts are listed in Table~\ref{tab:eventrate}. 

\renewcommand{\arraystretch}{1.3}
\begin{table}[h]
\centering
\caption{Expected number of CC and NC neutrino interaction events with the analysed detector volume, along with the uncertainty from the neutrino flux.
}
\vspace*{-3mm}
\small
\begin{tabular}{cccc}
\\
\hline
$\nu_e+\bar\nu_e$ CC & $\nu_\mu+\bar\nu_\mu$ CC & $\nu_\tau+\bar\nu_\tau$ CC & NC\\
\hline
8.5$^{+4.3}_{-1.8}$ & 43.6$^{+4.7}_{-5.2}$ & 0.12$^{+0.22}_{-0.05}$ & 16.5$^{+3.0}_{-2.1}$\\
\hline
\end{tabular}
\label{tab:eventrate}
\end{table}
\renewcommand{\arraystretch}{1}

\vspace{-2mm}
\section{Appendix C: Resolutions of electron energy and muon momentum measurements}

\begin{figure}[htb]
\centering 
\includegraphics[width=0.75\linewidth]{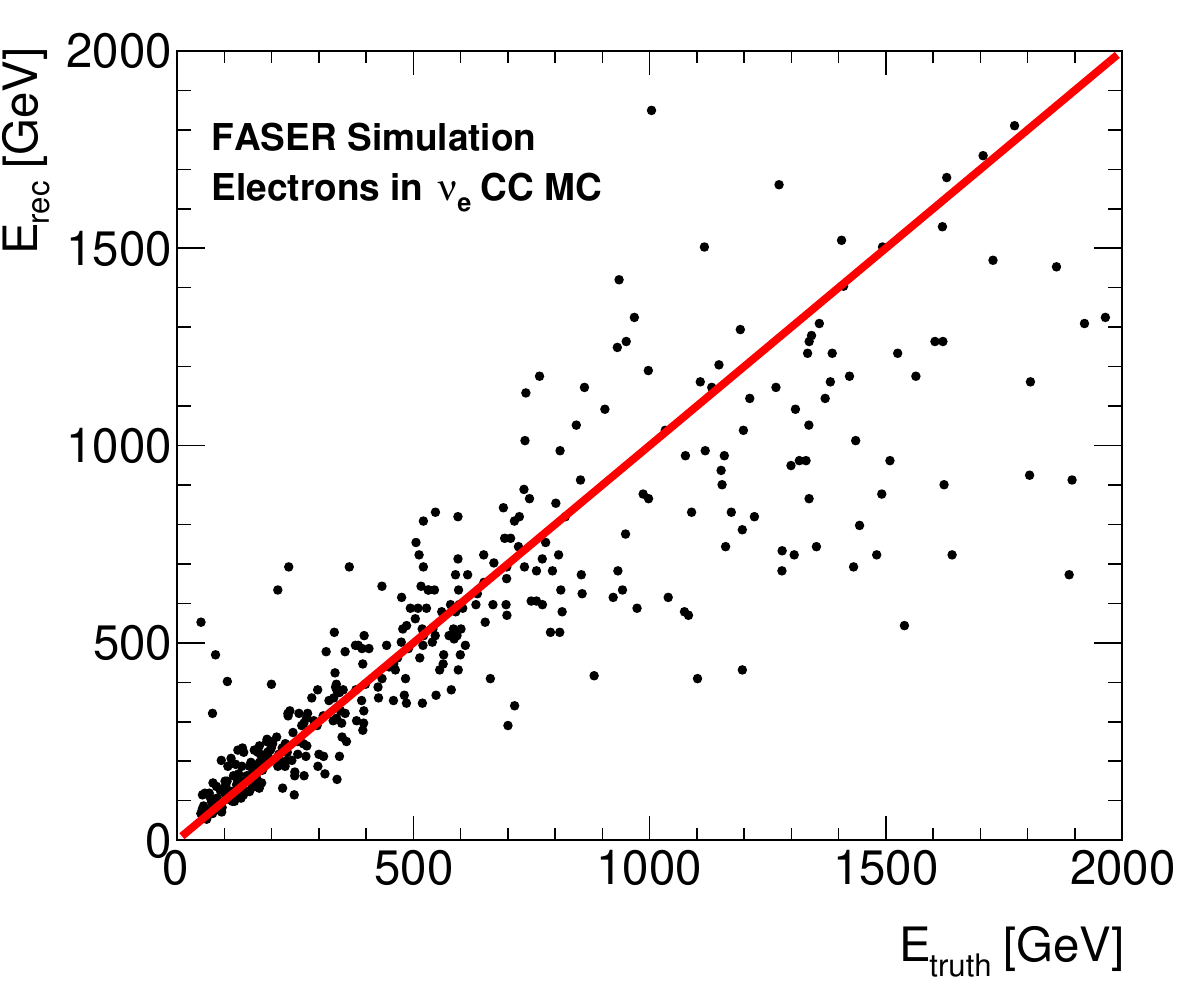}
\vspace*{-3mm}
\caption{The reconstructed electron energy versus the true energy in $\nu_{e}$ CC MC simulation. 
}
\label{fig:EMshower_Erec_Etruth}
\end{figure}

Appendix C describes the electron energy and muon momentum measurement resolutions used for the event selection. The energy reconstruction algorithm performance was tested for electrons in the $\nu_e$ MC simulation (Figure~\ref{fig:EMshower_Erec_Etruth}), showing a resolution of around 25\% at 200~GeV and between 25-40\% at higher energies. 

\begin{figure}[htb]
\centering
\includegraphics[width=0.8\linewidth]{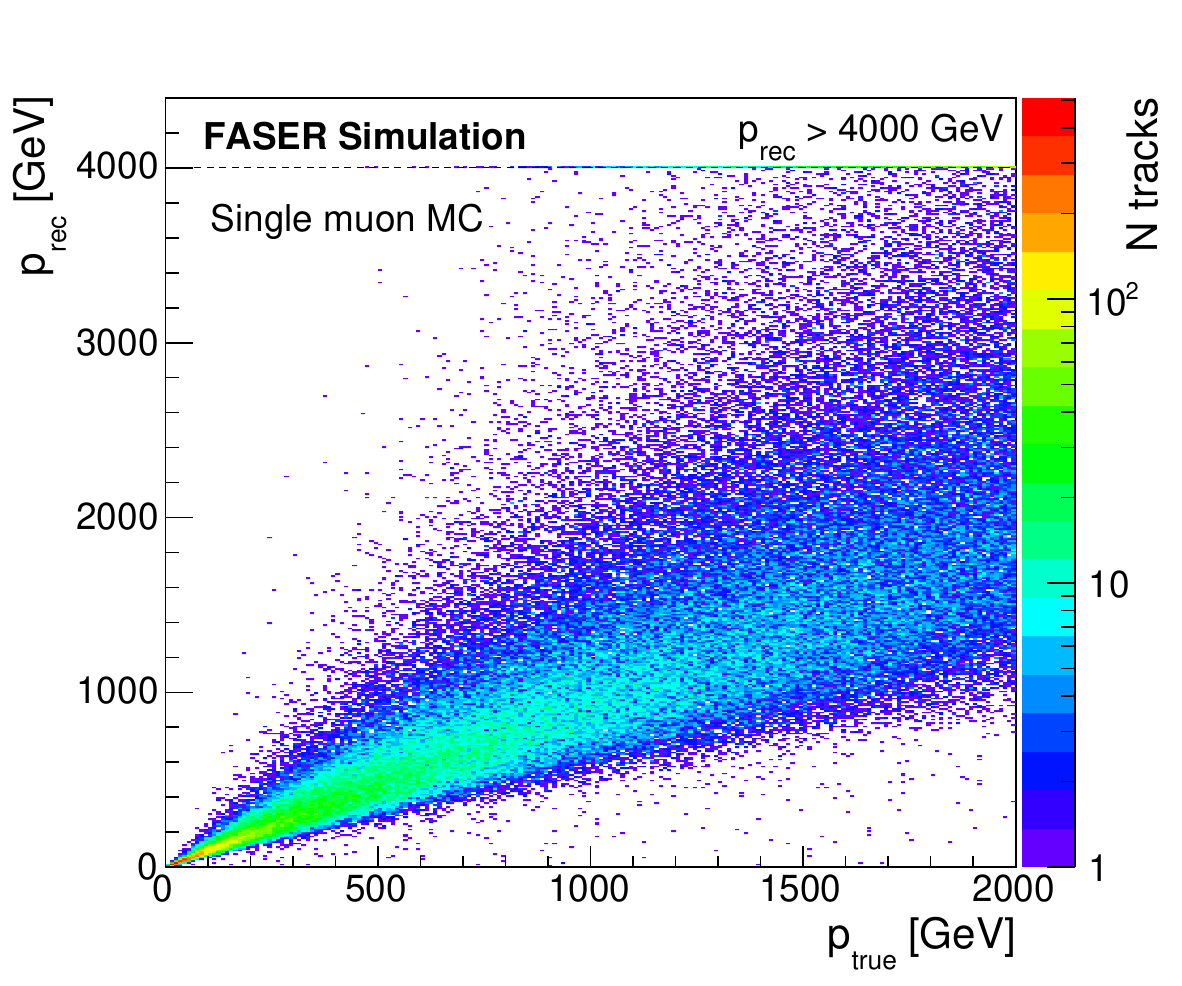}
\vspace*{-3mm}
\caption{Reconstructed momenta versus true momenta in simulated muon tracks with a flat momentum distribution from 1 to 2000 GeV.
}
\label{fig:ptrue_prec}
\end{figure}

The resolution, as quantified by the RMS of the distribution of the difference between the true momentum and reconstructed momentum, is around 30\% at 200~GeV and reaches 50\% at higher energies; see Figure~\ref{fig:ptrue_prec}. 
The track momentum assessment performance is validated with data by studying the momenta of long tracks: they are split into two tracks, and the reconstructed momenta of the two halves is compared, resulting in a reasonable agreement.

\vspace{-2mm}
\section{Appendix D: Validation of the modelling of the neutral-hadron background}
\begin{figure}[htb]
\centering
\includegraphics[width=0.95\linewidth]{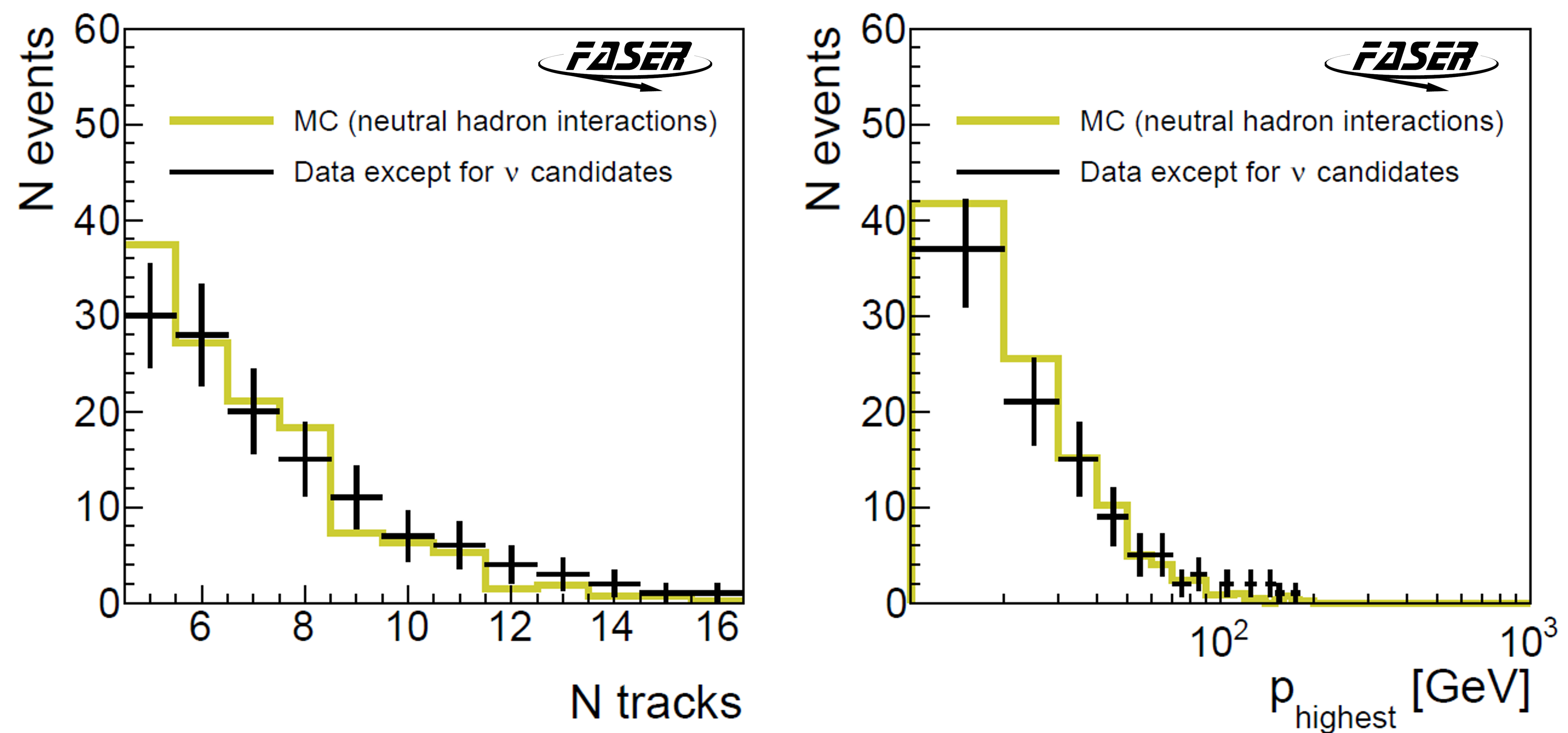}
\vspace*{-3mm}
\caption{MC simulation distributions of track multiplicity (left) and momentum of the highest momentum track (right) from neutral-hadron interactions vertices. The observed events in the data sample (except for neutrino candidate events) are shown in black. The MC simulation distributions are normalized to the number of events observed in the data with the normalization factor of 0.57.
}
\label{fig:plots_lowE_had_ev}
\end{figure}

Appendix D describes the validation of the modelling of the neutral-hadron background, using the initial neutral-vertex sample of data (before the high-energy electron or muon selection is applied), which is dominated by neutral-hadron interactions. For this validation study, only a part of the analysed volume (150 tungsten plates from film 7 to 156) was used. 
The expected number of hadron interaction vertices is 246, while the number of neutral vertices in the data sample is 139. 
Figure~\ref{fig:plots_lowE_had_ev} shows a comparison of the number of tracks in the vertex, and the reconstructed momentum of the highest momentum track, between the neutral-hadron MC and the data. For the comparison, the MC distributions are normalized to the same number of vertices as observed in the data. The neutrino candidates that satisfy the event selection are excluded from the data for this comparison. 
The distribution shapes are well modelled in the simulation, and the number of interactions is found to be compatible at better than the 50\% level, with more neutral-hadron interactions predicted in the MC than observed in the data.

\vspace{-2mm}
\section{Appendix E: Properties of the $\nu_{e}$ and $\nu_{\mu}$ CC candidate events}
\begin{figure}[h]
\centering
\includegraphics[width=0.77\linewidth]{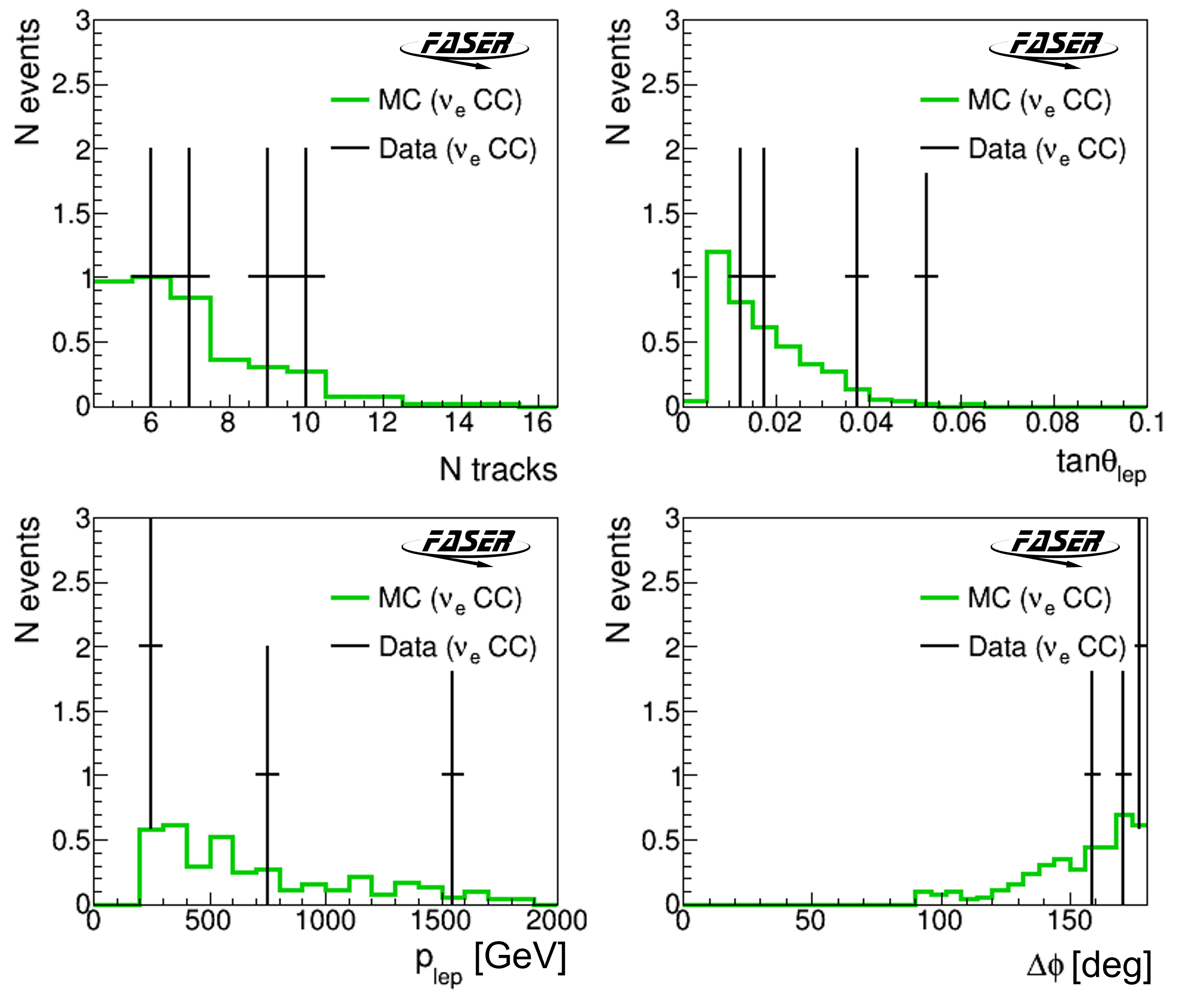}
\includegraphics[width=0.77\linewidth]{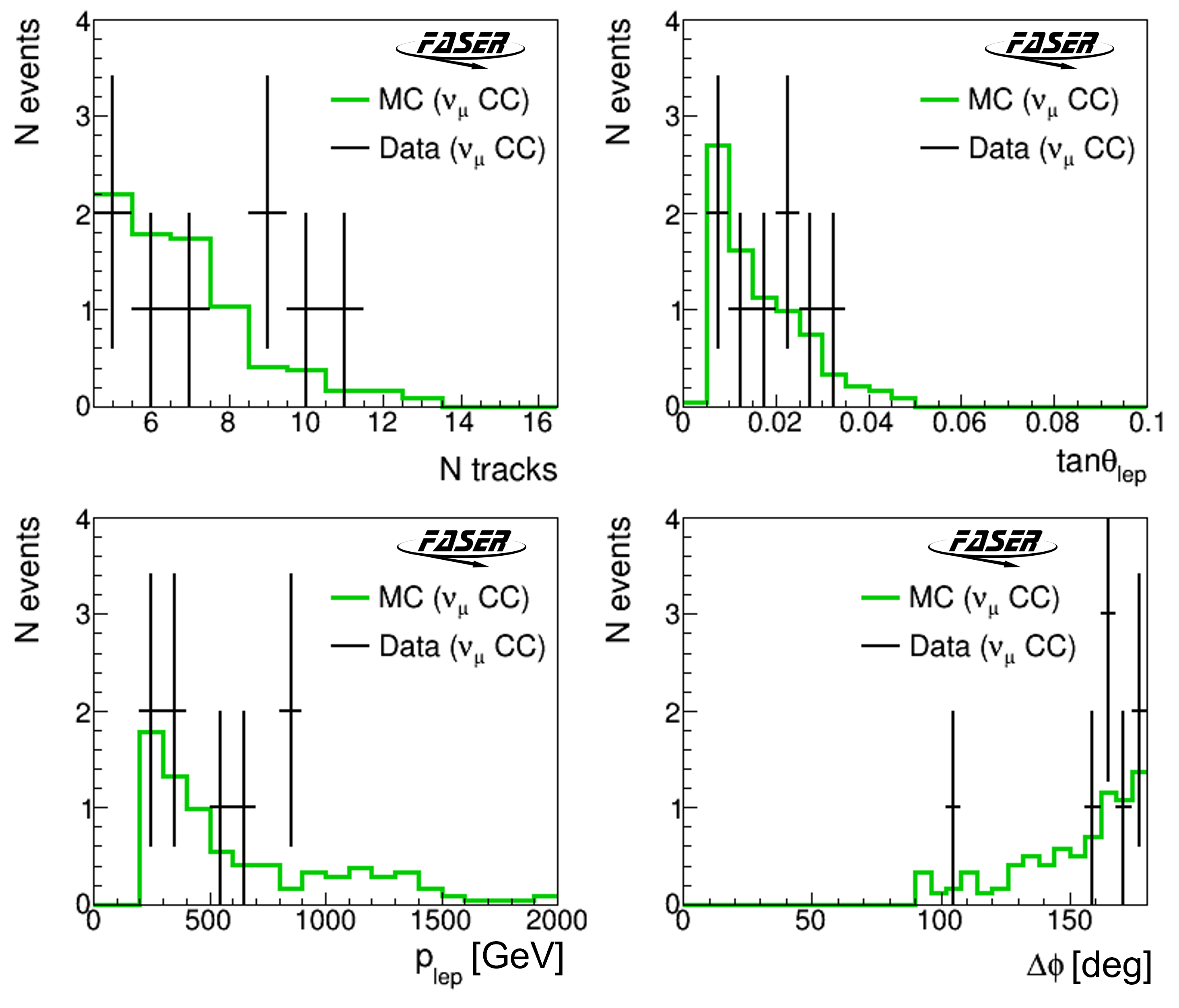}
\vspace*{-3mm}
\caption{MC simulation distributions of the track multiplicity (N tracks), lepton angle ($\tan \theta_{\text{lep}}$), lepton momentum ($p_{\text{lep}}$), and $\Delta\phi$ for the $\nu_{e}$ CC (top four figures) and $\nu_{\mu}$ CC (bottom four figures) signal that passed the selection criteria. The observed $\nu_{e}$ CC and $\nu_{\mu}$ CC candidate events in the data sample are shown in black. The MC simulation distributions are normalized to the number of observed events with the normalization factor of 2.3 (0.9) for $\nu_{e}$ ($\nu_{\mu}$).
}
\label{fig:plots_nuCC_ev}
\end{figure}

\begin{figure}[h]
\centering
\includegraphics[width=0.77\linewidth]{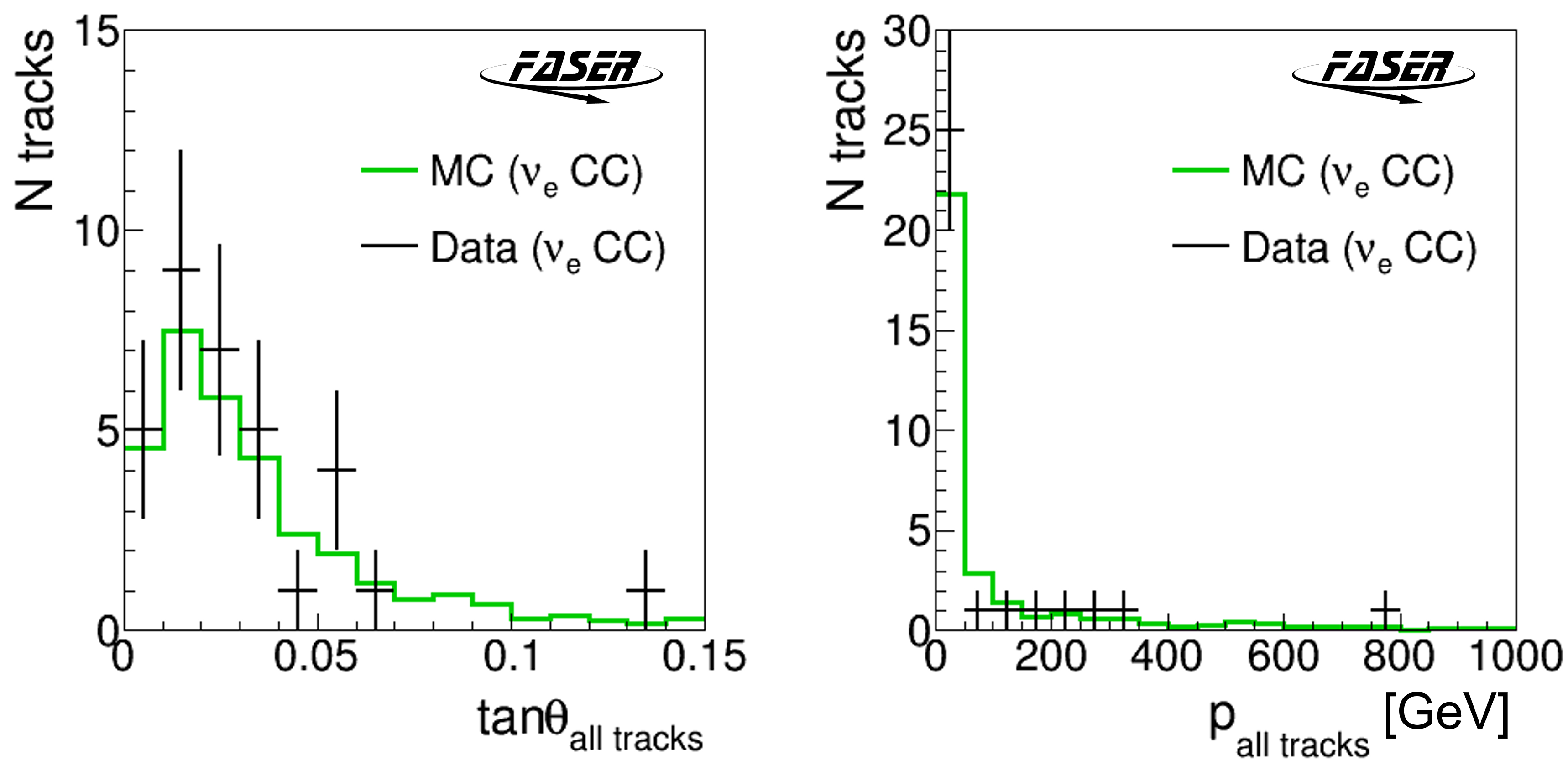}
\includegraphics[width=0.77\linewidth]{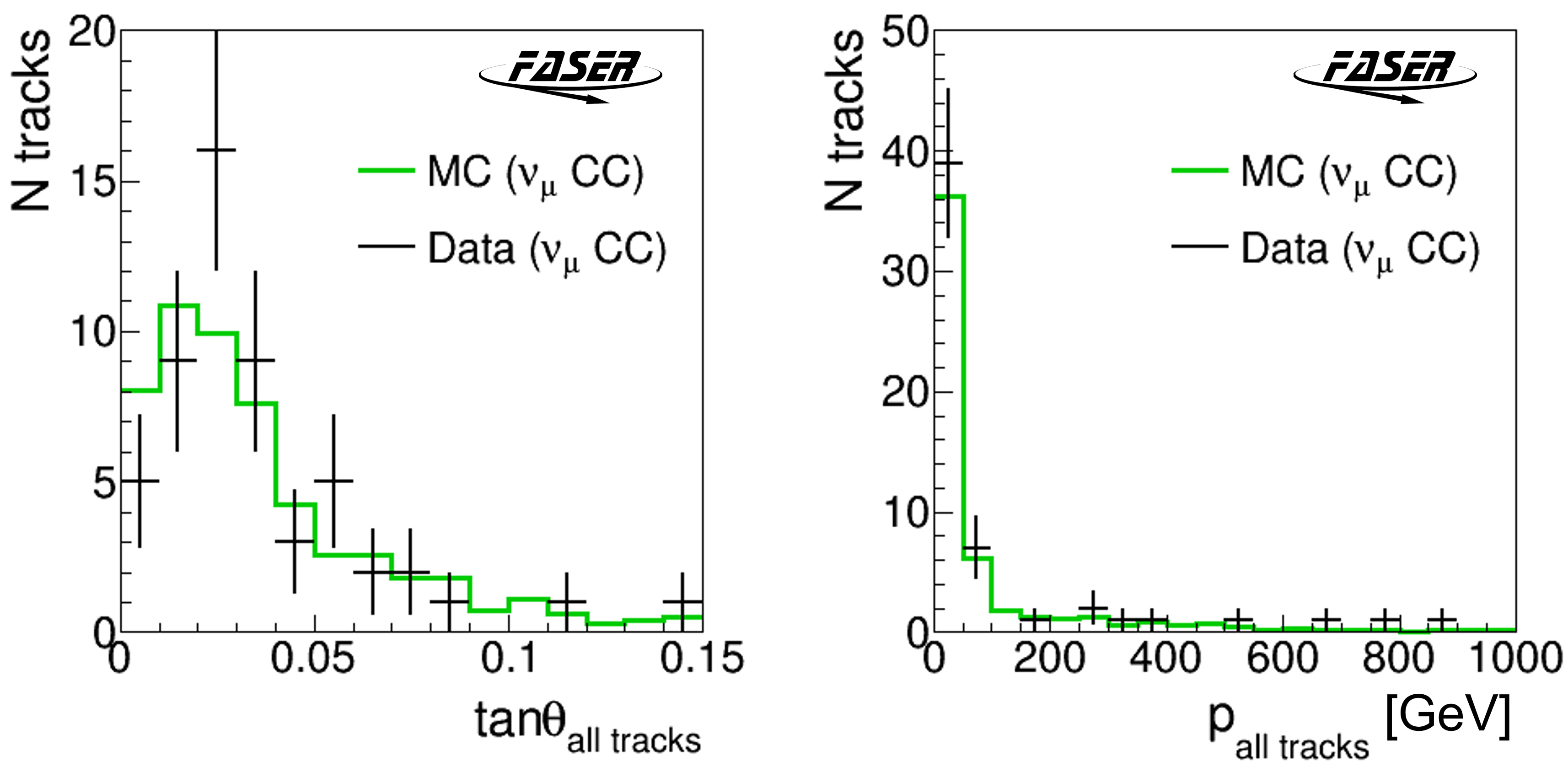}
\vspace*{-3mm}
\caption{
MC simulation distributions of track angle and momentum from vertices of the $\nu_e$ CC (top) and $\nu_{\mu}$ CC (bottom) signal. The observed $\nu_e$ CC and $\nu_{\mu}$ CC candidate events in the data sample are shown in black. The MC simulation distributions are normalized to the number of observed tracks.
}
\label{fig:plots_nuCC_trk}
\end{figure}

Appendix E describes the properties of the selected vertices compared with the expectations from $\nu_{e}$ CC and $\nu_{\mu}$ CC simulation (Figure~\ref{fig:plots_nuCC_ev}) and for the properties of the individual tracks forming the vertices (Figure~\ref{fig:plots_nuCC_trk}). In general the simulation describes the data well for both the $\nu_e$ and $\nu_\mu$ selections. 

\clearpage

\bibliographystyle{utphys}
\bibliography{main}

\end{document}